\documentclass{mem}
\usepackage{natbib}\usepackage{txfonts}\usepackage{balance}
\usepackage{graphicx}
\usepackage{txfonts}
\usepackage[a4paper]{hyperref}
\idline{79}{3}
\begin{document}
\def\teff{$T\rm_{eff }$}
\def\kms{$\mathrm {km s}^{-1}$}

\title{
Impact of Rotation on the Evolution
of Low--Mass Stars
}

   \subtitle{}

\author{
D. Brown\inst{1} 
\and M. Salaris\inst{1}
\and S. Cassisi\inst{2}
\and A. Pietrinferni\inst{2}
          }

  \offprints{D. Brown}

\institute{
Astrophysics Research Institute --
Liverpool John Moores University,
Twelve Quays House,
Egerton Wharf,
Birkenhead,
CH41 1LD,
United Kingdom\\
\email{db@astro.livjm.ac.uk}
\and
Instituto Nazionale di Astrofisica,
Osservatorio Astronomico di Collurania ``Vincenzo Cerulli'',
Via Mentore Maggini s.n.c.,
64100 Teramo,
Italy
}

\authorrunning{Brown and Salaris}

\titlerunning{Impact of Rotation on the Evolution
of Low--Mass Stars}

\abstract{
High precision photometry and spectroscopy of low--mass stars
reveal a variety of properties standard stellar evolution 
cannot predict. Rotation, an essential ingredient of stellar evolution,
is a step towards resolving the discrepancy between model predictions
and observations.

The first rotating stellar model, continuously tracing a
low--mass star from the pre--main sequence onto the horizontal branch,
is presented. The predicted luminosity 
functions of stars of globular clusters and surface rotation velocities
on the horizontal branch are discussed.

}
\maketitle{}
%%%%%%%%%%%%%%%%%%%%%%%%%%%%%%%%%%%%%%%%%%%%%%%%%%%%%%%
\section{Introduction}
Rotation has to be included in realistic stellar models.
Since recent high precision photometry and spectroscopy have become
available, the effects of slower rotation on low--mass stars has become
detectable in the colour--magnitude diagram (CMD) 
and Luminosity Functions (LFs).
To predict them precisely, low--mass stellar models
need to take rotation into account.
\subsection{Stellar Structure}
The version of the stellar evolution code FRANEC, described in \citet{Pietrinferni2004},
is used for our computations.
Rotation was included in the stellar structure equations
following the prescription of \citet{Kippenhahn1970}. 
The required parameters $f_{\rm P}$ and $f_{\rm T}$
were derived analytically from the effective gravitational potential
of the star approximating its equipotential surfaces as
rotational ellipsoids.

We have eliminated the subatmospheric region used in the
original code and increased the density of the outermost
meshpoints in the stellar model.
\subsection{Transport Mechanisms}
Chemical elements and angular momentum
are transported within the star during its evolution. 
In a convective zone (CZ) the 
transport is assumed to be instantaneous.
The abundance of chemical elements is averaged in a CZ and
the angular momentum redistributed to achieve either specific angular momentum
conservation or solid body rotation. Both distributions are limiting cases
introduced by \citet{Sweigart1970}. 
Solid body rotation is chosen for CZ 
until the the star reaches the turn off (TO), 
after which
the specific angular momentum is conserved
(see \citealt{Palacios2006b} and \citealt{Sills2000}).

In a radiative region the transport is a diffusive process,
since \citet{Palacios2006a} have shown that advective 
transport via meridional currents is less
efficient than via diffusion resulting from the shear instability.
This instability dominates rotational mixing \citep{Zahn1992}.
The diffusion coefficient $D_{v}$ is taken from \citet{Denissenkov2004} 
introducing an efficiency parameter $F$.
Chemical elements are transported by a combination of atomic diffusion 
and rotational mixing. Angular momentum is diffused by rotational mixing alone.
\subsection{Braking}
The total amount of angular momentum of the star is changed by several
processes. On the pre-main sequence (PMS) 
magnetic fields connect the star with its circumstellar disk. The disk allows
angular momentum to be diffused beyond the influence of the magnetic field of
the star, preventing the PMS spin-up. This process of
disk braking \citep{Shu1994} is active until a time $\tau_{\rm DB}$.

On the main sequence (MS) the stellar magnetic field 
increases the amount of angular momentum lost by its stellar wind.
The angular momentum lost by this magnetic--wind braking  
is given by \citet{Chaboyer1995}.
We include a function determining the efficiency to generate
a magnetic field $B_{\rm eff}$ and  parametrise it by the mass of the
convective envelope. 
The shape of $B_{\rm eff}$ is determined by 
two parameters ($M_w$;$B_w$). The overall efficiency of the braking process
is determined by $f_k$.

Angular momentum lost via mass loss is already included in the
code, and is dominant on the red giant branch (RGB).
%
%%%%%%%%%%%%%%%%%%%%%%%%%%%%%%%%%%%%%%%%%%%%%%%%%%%%%%%%
\section{Calibration}
The modified stellar evolution code 
contains a number of free parameters. First,
the two parameters describing the initial conditions of the star
have to be determined: the
initial rotation rate $\omega_0$ and the disk braking time $\tau_B$. 
Four additional parameters to be calibrated are:
the mixing length
$\alpha$, the braking efficiency $f_{\rm k}$, the function describing the
efficiency to generate the magnetic field $B_{\rm eff}$, and
the rotational mixing efficiency $F$.
\subsection{Initial Conditions} 
To determine $\omega_0$ we used the rotation period of a sample
of $\sim$1\,Myr old stars from three open cluster:
IC 348 \citep{Littlefair2005}, Orion Nebula Cluster \citep{Herbst2002}, 
and the Orion Nebula Cluster Flanking fields \citep{Rebull2001}.
We limited the mass range to $0.5\,M_\odot < M < 1.3\,M_\odot$ 
to later reproduce the surface rotation velocity for the Hyades.
The mean value of the rotation period at 1\,Myr is 
$P_{\rm 1\,Myr}=(5\pm 4)$\,d.
Appropriate values of $\omega_0$ for our stellar models
were chosen to reproduce $P_{\rm 1\,Myr}$.
This $\omega_0$ is derived from stars
with solar metallicity. To determine $\omega_0$
at lower metallicites, the total angular momentum of stars of 
the same mass, but different metallicity, is assumed to be the same at the start
of their life on the PMS.

The disk braking time is set to $\tau_{\rm DB}=5$\,Myr, at which 50\,\% of
the stars in clusters have no longer a circumstellar disk \citep{Bouvier1997}.
\subsection{Free Parameters}
\begin{figure}[]
\resizebox{\hsize}{!}{\includegraphics[clip=true]{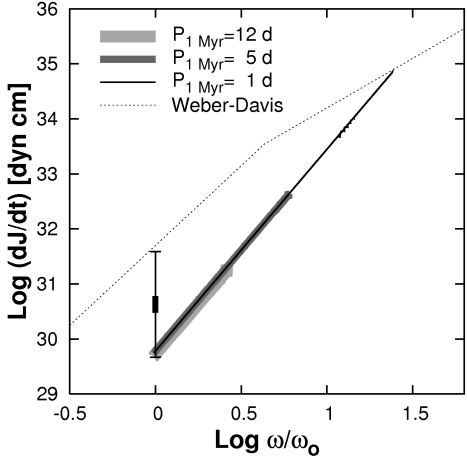}}
\caption{
\footnotesize
            Rate of angular momentum loss
            ($\frac{{\rm d}J}{{\rm d}t}$) by
            magnetic wind braking plotted against
            surface rotation rate of three different solar models
            ($P_{\rm 1\,Myr}$=5\,d grey line,
            $P_{\rm 1\,Myr}$=12\,d thick light grey line, and
            $P_{\rm 1\,Myr}$=1\,d thin black line). 
            Observed 
            $\frac{{\rm d}J}{{\rm d}t}$ values for the Sun  
            ($\log \frac{\omega}{\omega_0}$=0) and a
            theoretically derived upper limit (dotted line) for the MS is shown
            (see text for details).
}
\label{weberdavis}
\end{figure}

To calibrate the mixing length ($\alpha$) and $f_k$ we created
a theoretical solar model.
The predicted rate of angular momentum loss 
$\left(\frac{{\rm d}J}{{\rm d}t}\right)$ for the
model was used to confirm the adopted braking mechanisms.
In Fig.~\ref{weberdavis} we give the rate of angular momentum loss for three
different solar models with initial rotation rates consistent with 
observed $P_{\rm 1\,Myr}$ ($P_{\rm 1\,Myr}$=5\,d grey line,
$P_{\rm 1\,Myr}$=12\,d thick light grey line, and  $P_{\rm 1\,Myr}$=1\,d
thin black line). Each of these models reaches $\left(\frac{{\rm d}J}{{\rm d}t}\right)$
observed for the Sun and shown in Fig.~\ref{weberdavis}. 
Additionally, they lie below a
theoretical upper limit determined by a 2D \citet{Weber1967} equatorial wind model. 

\begin{figure}[]
\resizebox{\hsize}{!}{\includegraphics[clip=true]{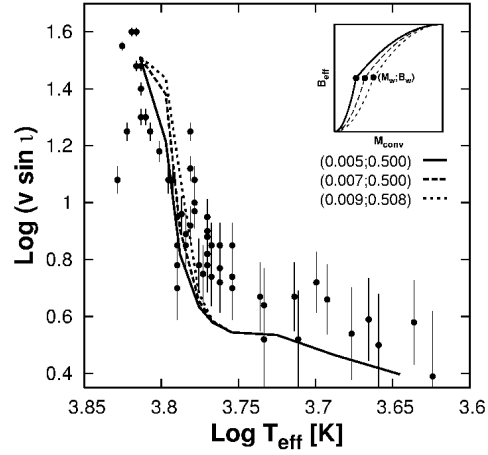}}
\caption{
\footnotesize
            The observed $\langle v \sin \iota \rangle$
            taken from \citet{Soderblom1993}
            as a function of 
            T$_{\rm eff}$.
            The profile of the $B_{\rm eff}$ functions are given
            in the inset to the upper right, including the point ($M_{\rm w}$;$B_{\rm w}$)
            that defines its shape. Isochrones
            for $\langle v \sin \iota \rangle$ are shown using different
            $B_{\rm eff}$ functions:
            (0.005;0.500) solid, 
            (0.007;0.500) long dashed, and (0.009;0.508) 
            short--dashed line.}
\label{Hyades}
\end{figure}

$B_{\rm eff}$ was determined
by predicting surface rotation velocities for the Hyades and comparing them
with observations by \citet{Soderblom1993}. Figure
\ref{Hyades} shows three predicted rotation distributions for different
choices of the $B_{\rm eff}$ function ( (0.005;0.500) solid, 
(0.007;0.500) long dashed, and (0.009;0.508) short--dashed line).
The profile of the $B_{\rm eff}$ functions are given
in the inset to the upper right, including the point ($M_{\rm w}$;$B_{\rm w}$)
that defines its shape.
The best fit between data and models was reached with (0.005;0.500). 

Spectroscopic observations that in the past were explained in terms of
efficient rotational mixing are now being challenged by new
data and theoretical developments (see Spite plateau
data of \citealt{Asplund2006} presented in \citealt{Bonifacio2007};
\citet{Eggleton2007}, P. Ventura and E. Carretta in this proceedings,
for the issue of extra mixing on the RGB).
Therefore, the rotational mixing efficiency $F$ is set to an
arbitrary low value of $10^{-6}$, allowing for a marginal transport  
without creating any significant
changes to the chemical abundances in the stellar model.
%
%%%%%%%%%%%%%%%%%%%%%%%%%%%%%%%%%%%%%%%%%%%%%%%%%%%%%%%
\section{Application to GC}
Using our rotating stellar models we present a comprehensive analysis
of rotational effects on the evolution of low--mass stars in globular
clusters. Past work was exploratory \citep{Vandenberg1998} or
targeted specific issues \citep{Palacios2006a}.
We apply our models to typical globular cluster
stars (Z=$10^{-3}$ and Y=0.246) and corresponding isochrones
of 10.5 and 12.0\,Gyr.
\subsection{Surface Velocities}
Our isochrones predict a
surface velocities on the SGB 
$v_{\rm surf}=1-5.5$\,km\,s$^{-1}$, consistent with the observed
$v_{\rm surf}\sim 3$\,km\,s$^{-1}$ by \citet{Lucatello2003}. More
precise observations of the rotation along the SGB
are necessary to further constrain the initial conditions
and the angular momentum transport in CZs.

Rotational velocities on the HB were determined
for a 0.8\,M$_\odot$ star experiencing no mass loss with an optimised version 
of our rotational code that is able to evolve through
the helium flash. We predicted 
$v_{\rm surf}=6.85$\,km\,s$^{-1}$ on the zero--age HB (ZAHB)
that is within the range observed for
red HB stars \citep{Sills2000}.
\subsection{Isochrones}
\begin{figure}[]
\resizebox{\hsize}{!}{\includegraphics[clip=true]{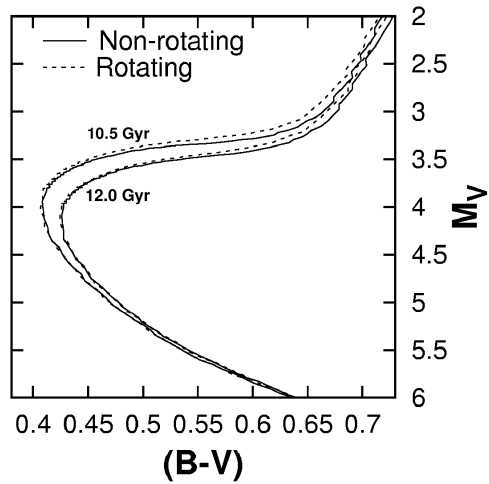}}
\caption{
\footnotesize
            The 10.5\,Gyr and 12\,Gyr
            isochrones without rotation (as labelled)
            are plotted in a BV CMD as solid lines. The
            isochrones including rotation are shown as dashed lines.}
\label{iso}
\end{figure}

Two isochrones for ages of 10.5 and 12.0\,Gyr
are shown in Fig.~\ref{iso}, with (dashed line)
and without rotation (solid line). 
Rotation does not shift the MS or the TO,
however,
the brightness of the SGB at (B-V)=0.55 is slightly increased 
by 0.05 and 0.03\,mag, for 10.5 and 12.0\,Gyr respectively. The
RGB at $M_V=2$ including rotation is bluer by 0.01 and 0.007\,mag
for 10.5 and 12.0\,Gyr respectively.

Rotation increases
the tip of the RGB (TRGB) luminosity by
$\Delta\log{L/L_\odot}=0.08$ and the helium core mass at the
flash by $\Delta M_{\rm He}=0.048$\,$M_\odot$. This large helium core
leads to a brighter ZAHB luminosity of $\Delta M_V=0.326$ 
when including rotation, and will be discussed later.

\subsection{Luminosity Functions}
From the isochrones we then determine the LF for the
case of rotation and no rotation
using a \citet{Salpeter2004} initial mass function. 
Figure \ref{lf} shows the two cases are normalised
at a $M_V$=4 which is close to the TO and independent of the choice
of initial mass function. The shape of the RGB bump is not changed 
by rotation, only its location is shifted to brighter luminosities
by $\Delta M_V=0.1$ for the 10.5\,Gyr LF. Similar to
\citet{Vandenberg1998}, we find that rotation reduces the SGB slope and
increases the number count at the base of the RGB (by 11\,\% for the
12\,Gyr LF). 
%
%%%%%%%%%%%%%%%%%%%%%%%%%%%%%%%%%%%%%%%%%%%%%%%%%%%% 
\section{Discussion}
Given an isochrone without rotation that is consistent with
observational constraints. The inclusion of
rotation increases the helium core mass, this leads to a brighter
ZAHB and distances from HB fitting that are no
longer consistent with parallax based
empirical MS fitting \citep{Recio2004}.

A solution would be to reduce the rotation rate in the 
helium core on the RGB. This reduction may result from either
more realistic angular momentum transport in CZs
(see exploratory work by
\citealt{Palacios2006b})
or reducing $\omega_0$. 

We chose to
reduce the initial rotation rate by adopting a different method
to transpose solar metallicity rotation rates to lower metallicities.
Instead of keeping the total amount of initial angular momentum constant,
$\omega_0$ is defined so that these stars reach a rotational period
of $P_{\rm 1\,Myr}$=5\,d. This reduces $\omega_0$ by 60\,\% and
leads to a SGB rotation velocity of 3\,km\,s$^{-1}$,
consistent with observations. The ZAHB including rotation
is only brighter by $\Delta M_V$=0.122
(comparable to typical photometric errors for HB stars) and leads to
ages younger by 1.5\,Gyr. Therefore, an isochrone or LF 
with and without rotation can only be consistent with 
observational constraints, if the rotational isochrone or LF is
1.5\,Gyr younger than without rotation. 

\begin{figure}[]
\resizebox{\hsize}{!}{\includegraphics[clip=true]{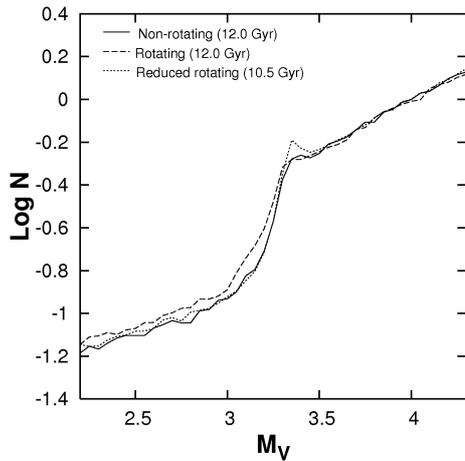}}
\caption{
\footnotesize
            Three LFs are shown as a function of
            $M_V$ with an age of 12\,Gyr without
            (solid line) and with rotation (dashed lines), as well
            as an age of 10.5\,Gyr including the reduced initial
            rotation (dotted line). 
}
\label{lf}
\end{figure}

A consistent comparison between a 10.5\,Gyr old LF with the reduced
$\omega_0$ (solid line) and a 12.0\,Gyr LF without (dashed line), 
is shown in Fig.~\ref{lf}. The prior determined 11\,\% increase
in number count is reduced to 5\,\%, but it is still detectable.
%%%%%%%%%%%%%%%%%%%%%%%%%%%%%%%%%%%%%%%%%%%%%%%%%%%%%%%%%%
\begin{acknowledgements}
I am grateful to S. Cassisi and A. Pietrinferni for enabling
several research stays at Teramo observatory for me and the
valuable discussions furthering the understanding of FRANEC and
the helium flash code.
\end{acknowledgements}
%
%%%%%%%%%%%%%%%%%%%%%%%%%%%%%%%%%%%%%%%%%%%%%%%%%%%%%%%%%

\bibliographystyle{aa}

\end{document}